\documentclass[aps,pra,a4,superscriptaddress,preprintnumbers,twoside,twocolumn,floatfix]{revtex4}
\usepackage{stmaryrd}
\usepackage{graphicx}
\usepackage{color}
\usepackage{bm,bbold}
\usepackage{bbm}
\usepackage{enumerate}
\usepackage{color}
\graphicspath{figures} 
\usepackage{amsfonts, amsmath, amsthm, amssymb} 
\usepackage{mathtools}
\usepackage{dsfont}
\usepackage{array}
\usepackage{changes}

\usepackage[colorlinks,bookmarks=false,citecolor=blue,linkcolor=red,urlcolor=blue]{hyperref}
\begin{document}

\title{Optimal entanglement witnesses: a scalable data-driven approach}

\author{Ir\'en\'ee Fr\'erot}
\email{irenee.frerot@gmail.com}
\affiliation{ICFO-Institut de Ciencies Fotoniques, The Barcelona Institute of Science and Technology, Av. Carl Friedrich Gauss 3, 08860 Barcelona, Spain
}
\affiliation{Max-Planck-Institut f{\"u}r Quantenoptik, D-85748 Garching, Germany}
\author{Tommaso Roscilde}\email{tommaso.roscilde@ens-lyon.fr}
\affiliation{Laboratoire de Physique, CNRS UMR 5672, Ecole Normale Sup\'erieure de Lyon, Universit\'e de Lyon, 46 All\'ee d'Italie, Lyon, F-69364, France
}

\begin{abstract}
Multipartite entanglement is a key resource allowing quantum devices to outperform their classical counterparts, and entanglement certification is fundamental to assess any quantum advantage.  The only scalable certification scheme relies on entanglement witnessing, typically effective only for special entangled states. Here we focus on finite sets of measurements on quantum states (hereafter called quantum data); and we propose an approach which, given a particular spatial partitioning of the system of interest, can effectively ascertain whether or not the data set is compatible with a separable state. When compatibility is disproven, the approach produces the optimal entanglement witness for the quantum data at hand. 
Our approach is based on mapping separable states onto equilibrium classical field theories on a lattice; and on mapping the compatibility problem onto an inverse statistical problem, whose solution is reached in polynomial time whenever the classical field theory does not describe a glassy system. Our results pave the way for systematic entanglement certification in quantum devices, optimized with respect to the accessible observables. 
\end{abstract}

\maketitle

\noindent \textit{Introduction.} Preparing and processing strongly entangled many-body states, in both a controlled and scalable way, is the goal of all quantum simulators and computers. Indeed, as the efficient representation of generic entangled many-body states is impossible on classical machines, entanglement represents the key computational resource of these devices \cite{georgescuetal2014,preskill2012quantum}. As a consequence, developing generic and scalable methods to certify entanglement in multipartite systems stands as a grand challenge of quantum information science. Even more fundamentally, entanglement certification is a central task to probe the resilience of quantum correlations from the microscopic world to the macroscopic one \cite{sangouard2018}.  

\begin{figure*}
    \centering
    \includegraphics[width=0.95\textwidth]{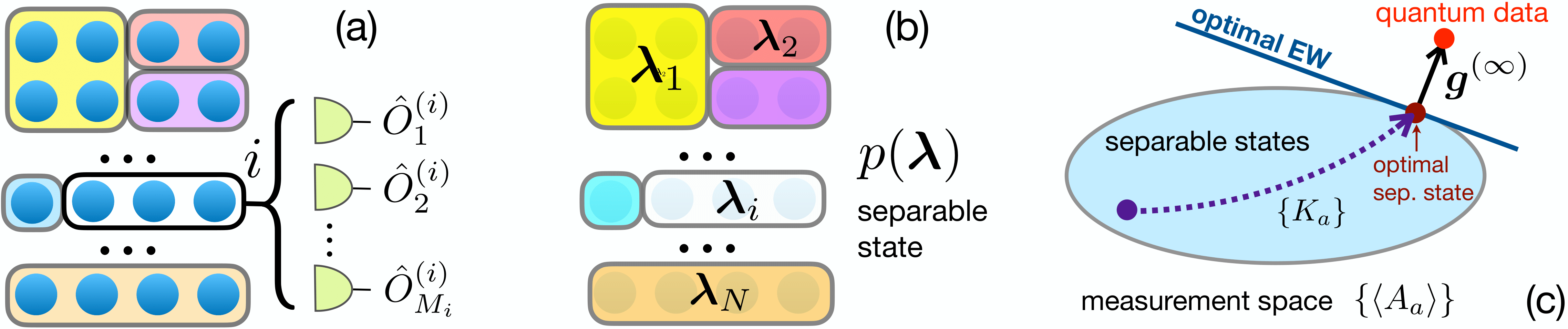}
    \caption{(a) Partition of a quantum device into $N$ clusters, each of which is subject to $M_i$ measurements; (b) A separable state of the system is described as a probability distribution $p(\bm \lambda)$ of local states defined by the $\{ \bm \lambda_i \}$ parameters; (c) Our algorithm builds a trajectory of separable states (parametrized by couplings $\{K_a\}$ defining $p(\bm \lambda)$) which converges to the optimal state approximating at best some target quantum data. If the state fails to reproduce the quantum data exactly, the vector joining the optimal separable data and the quantum data reconstructs the optimal EW inequality.}
    \label{fig_schema_entanglement}
\end{figure*}

Any practical method must circumvent the tomographic reconstruction of the full density matrix \cite{paris2004quantum,dohertyetal2005} (which implies a number of measurements scaling exponentially with system size), and it should instead infer entanglement from the partial information contained in a given data set of measurement results (hereafter referred to as \emph{quantum data}). When one adopts this data-driven strategy, the goal of entanglement certification is to establish whether or not the quantum data are compatible with a separable state \cite{dohertyetal2005,Guhneetal2008,navascues2020entanglement}.
Given an extended quantum system composed of $N_{\rm tot}$ degrees of freedom, grouped together into $N \leq N_{\rm tot}$ clusters [see Fig.~\ref{fig_schema_entanglement}(a)], the state $\hat\rho$ of the system is separable \cite{werner1989} if it can be written in the form
  \begin{equation}
	\hat\rho_p := \int d \bm \lambda ~ p(\bm \lambda)~ \hat\rho_{\rm prod}(\bm \lambda) 
	\label{eq_separable_states}
\end{equation}
where  $\hat\rho_{\rm prod}(\bm \lambda)  = \otimes_{i=1}^N |\psi_i(\bm \lambda_i)\rangle \langle \psi_i(\bm \lambda_i)| $ is a product state of the partition, $|\psi_i(\bm \lambda_i)\rangle$ being the state of the $i$-th cluster, parametrized by parameters $\bm \lambda = (\bm \lambda_1, ..., \bm \lambda_i,..., \bm \lambda_N)$, distributed according to $p(\bm \lambda)\geq 0$. The distribution $p$ fully specifies classical correlations across the partition. A multipartite entangled state $\hat\rho$, on the other hand, cannot be written in the above form. 
Given a set of observables $\hat A_a$ $(a = 1, ..., R)$, multipartite entanglement is therefore \emph{witnessed} by the quantum data set $\{\langle \hat A_a \rangle_{\hat\rho}\}_{a=1}^R$ [where $\langle \hat A_a \rangle_{\hat\rho} = {\rm Tr}(\hat A_a \hat\rho)$] if one proves that the latter cannot be reproduced by \emph{any} separable state. This task is accomplished by proving that the quantum data violate an entanglement witness (EW) inequality, $\langle \hat{\cal W} \rangle_{\hat\rho_p} = \sum_a W_a \langle \hat A_a \rangle_{\hat\rho_p} \geq B_{\rm sep}$, valid for all separable states $\hat\rho_p$ \cite{guhneT2009}. Here $W_a$ are suitable coefficients and $B_{\rm sep}$ is the so-called separable bound. 

EW operators $\hat{\cal W}$ are generally defined based on the properties of special entangled states (e.g. squeezed states, total spin singlets, etc.) \cite{guhneT2009}, and failure of a data set to violate a given EW inequality does not exclude the existence of a different violated inequality involving the same data, yet to be discovered. This may erroneously suggest that entanglement witnessing is limited by creativity and physical insight; and that the entanglement witnessing problem (``is a quantum data set compatible with a separable state?") \cite{dohertyetal2005,Guhneetal2008,navascues2020entanglement} is generically undecidable. The goal of our work is to show that this is \emph{not} the case, and that the entanglement witnessing capability of a quantum data set can be \emph{exhaustively} tested. Our key insight is that the problem of finding the distribution $p(\bm \lambda)$, which defines the separable state reproducing at best the quantum data, is a statistical inference problem; and remarkably it has the structure of a convex optimization problem, whose solution can be attained in a time scaling polynomially with the partition size (under mild assumptions), and with the Hilbert space dimension of the subsystems composing the partition. When the optimal separable state fails to reproduce the quantum data, the distance between the quantum data set $\{ \langle \hat A_a \rangle_{\hat\rho} \}$ and the optimal separable set  $\{ \langle \hat A_a \rangle_{\hat\rho_p} \}$ allows one to reconstruct the optimal EW inequality violated by the quantum data.  
 We benchmark our approach by establishing new EW inequalities satisfied by the low-temperature states of the Heisenberg antiferromagnetic chain and the quantum Ising chain; in the latter case, our new EW inequalities outperform all previously known EW criteria for multipartite entanglement. Our work parallels the recent mapping of the Bell-nonlocality detection problem onto an inverse statistical problem \cite{frerotR2020}, and it offers an efficient scheme for entanglement detection in state-of-the-art quantum devices within a device-dependent scenario. 
 
\noindent \emph{Quantum data set.} For definiteness, we assume that on each subsystem $i =1,..., N$, $M_i$ local observables $\hat O_m^{(i)}$ can be measured ($m=1, \ldots M_i$; 
\emph{e.g.}~the Pauli matrices $\hat\sigma_a^{(i)}$, $a\in\{x,y,z\}$ for individual qubits taken as subsystems). 
For convenience, we denote the local identity operator by $\hat O_0^{(i)} := \mathbb{1}$. In order to reveal entanglement, these local observables must be non-commuting ($[\hat O_m^{(i)}, \hat O_{n}^{(i)}] \neq 0$ for $1 \le m<n\le M_i$) \cite{footnote1}. From these local observables, we build $p$-body correlators of the form $\hat O_{\bm m} = \otimes_{i=1}^N \hat O_{m_i}^{(i)}$ where $m_i=0$ for $N-p$ subsystems. Arbitrary observables can be built as linear combinations of correlators --  such as \emph{e.g.}~powers of collective spin variables \cite{tothetal2009,vitaglianoetal2011} $\hat J_a = \sum_i \hat\sigma_a^{(i)}/2$ ($a=x,y,z$) for systems of qubits.
Hence we shall assume that $R$ observables of the form $\hat A_a = \sum_{\bm m} x_{\bm m}^{(a)} \hat O_{\bm m}$ can be measured, where the sum runs over all strings ${\bm m}=(m_1, \ldots m_N)$, and $x_{\bm m}^{(a)}$ are arbitrary real coefficients. The quantum data $\{\langle \hat A_a \rangle_{\hat\rho}\}_{a=1}^R$ form the basis for entanglement certification in our scheme.
The problem of entanglement certification based on a data set has been discussed in the past, but the proposed methods either lack scalability \cite{Guhneetal2008}, or are scalable only under some restrictive assumptions (short-range correlations, low-dimensional geometry) \cite{navascues2020entanglement}. Our method aims at surpassing these limitations. 

\noindent \textit{Mapping onto an inverse statistical problem.}
 The key aspect behind our approach is the limited information content of separable states. The parameters $\bm \lambda$ specifying the product state $\hat\rho_{\rm prod}(\bm \lambda)$ can indeed be chosen as $\sum_i (2 d_i-2) \sim {\cal O}(N)$ real parameters, where $d_i$ is the dimension of the local Hilbert space of the $i$-th subsystem \cite{footnote2}.  
 The average of the $\hat A_a$ observable on a separable state reads
 \begin{equation}
\langle \hat A_a\rangle_{\hat\rho_p} = \int d\bm\lambda ~p(\bm \lambda) {\cal A}_a(\bm \lambda) =: \langle {\cal A}_a \rangle_p
\label{e.Ap}
 \end{equation}
 where ${\cal A}_a(\bm \lambda) = \sum_{\bm m} x_{\bm m}^{(a)} \prod_{i=1}^N o_{m_i}^{(i)}(\bm \lambda_i)$ and $o_{m_i}^{(i)}(\bm \lambda_i) = \langle \psi_i(\bm \lambda_i) | \hat O_{m_i}^{(i)} | \psi_i(\bm \lambda_i) \rangle$. Given a product state, the calculation of each term in the sum defining ${\cal A}_a(\bm \lambda)$ is clearly an operation scaling as ${\cal O}(N)$. 
Once the quantum nature of the state has been absorbed in ${\cal A}_a(\bm \lambda)$, the calculation of $\langle \hat A_a\rangle_{\hat\rho_p}$, Eq.~\eqref{e.Ap}, is a classical statistical average over the distribution $p$ which, from a statistical physics viewpoint, can be regarded as the Boltzmann distribution $p(\bm \lambda) =: \exp[-H(\bm \lambda)]/{\cal Z}$ of a classical field theory on a lattice (normalized by the $\cal Z$ factor), with a vector field $\bm \lambda_i$ defined on each of the $N$ clusters [Fig.~\ref{fig_schema_entanglement}(b)]. The complexity of separable states is fundamentally inscribed in the effective Hamiltonian $H(\bm \lambda)$, which is \emph{a priori} arbitrary, namely it is specified by a number ${\cal O}(\exp(N))$ of parameters. 
 
 Once the classical statistical structure of the expectation values on separable states is exposed, the problem of reproducing the quantum data with a separable state takes the form of a statistical inference problem, whose solution is well known in statistical physics \cite{nguyen_inverse_2017}. First of all, applying a maximum-entropy principle \cite{jaynes1957}, the Hamiltonian can be parametrized without loss of generality with as many parameters as the elements of the quantum data set \cite{SM}:
 \begin{equation}
 H(\bm \lambda) = -\sum_{a=1}^R K_a {\cal A}_a(\bm \lambda)~.
 \end{equation}
 The parameters ${\bm K} = \{K_a\}_{a=1}^R$ -- the coupling constants of the classical field theory -- are Lagrange multipliers whose optimization allows one to build the separable state $\hat\rho_p$ whose expectation values $\{ \langle \hat A_a \rangle_{\hat\rho_p} \}$ best approximate the quantum data $\{ \langle \hat A_a \rangle_{\hat\rho} \}$. 
 The optimization of ${\bm K}$ can be efficiently achieved upon minimizing the cost function $L({\bm K}) := \log {\cal Z}({\bm K}) - \sum_a K_a \langle \hat A_a \rangle_{\hat\rho}$ \cite{frerotR2020,nguyen_inverse_2017}. The $a$-th component of the gradient of $L$ is $g_a := \frac{\partial L}{\partial K_a} = \langle {\cal A}_a \rangle_{p} - \langle \hat A_a \rangle_{\hat\rho}$, and its Hessian matrix is $\frac{\partial^2 L}{\partial K_a \partial K_b} = \langle {\cal A}_a {\cal A}_b \rangle_p - \langle {\cal A}_a \rangle_p \langle {\cal A}_b \rangle_p$, namely the covariance matrix of the ${\cal A}_a(\bm \lambda)$ functions. Since the latter is a semi-definite positive matrix, $L$ is a convex function. Therefore, a simple gradient-descent algorithm, which consists in iterating the update rule $K_a' = K_a - \epsilon[\langle {\cal A}_a \rangle_{p} - \langle \hat A_a \rangle_{\hat\rho}]$ with $\epsilon \ll 1$, or any improvement thereof, is guaranteed to reach the global optimum of the problem. In practice, this requires to repeatedly compute $\langle {\cal A}_a \rangle_{p}$ as in Eq.~\eqref{e.Ap}, a task efficiently accomplished \emph{e.g.} by Markov-chain Monte Carlo sampling of $p(\bm \lambda)$, whenever the Hamiltonian $H$ does not describe a glassy system.
 The restriction to non-glassy systems is the only practical limitation of our approach \cite{SM}; and is ensured in the examples considered below by considering translationally invariant systems. 

\noindent \textit{Construction of an optimal entanglement witness.} As illustrated on Fig.~\ref{fig_schema_entanglement}(c), the algorithm converges to the distribution $p$ which minimizes $|{\bm g}|$ -- the norm of the gradient of $L$. 
If the minimal distance $g^{(\min)}$ vanishes (within the error on the quantum data), i.e. if $\langle \hat A_a \rangle_{\hat\rho_p^{(\min)}} = \langle \hat A_a \rangle_{\hat\rho}$ for all $a=1, \ldots R$, then entanglement cannot be assessed from the available data. But in the opposite case, the coupling constants $K_a$ increase indefinitely along the optimization, and the coefficients of the gradient $g_a^{(\min)} = \langle \hat A_a \rangle_{\hat\rho_p^{(\min)}} - \langle \hat A_a \rangle_{\hat\rho}$ allow us to build a violated EW inequality. First, we define the normalized coefficients $W_a := -g^{(\min)}_a / |{\bm g}^{(\min)}|$. The condition $|{\bm g}^{(\min)}|^2 > 0$ is then rewritten as:
\begin{equation}
		-\sum_a W_a \langle \hat A_a \rangle_{\hat\rho}	  
	  <
	  \min_{\hat\rho_p} \left\{
		-\sum_a W_a \langle \hat A_a \rangle_{\hat\rho_p} 
	 \right\} =: B_{\rm sep}
	 	\label{eq_optimal_EW}
\end{equation}
The linear combination $\hat{\cal W} := -\sum_{a=1}^R W_a \hat A_a$ is the data-driven EW operator. The separable bound $B_{\rm sep}$, namely the minimal value of ${\rm Tr}(\hat\rho \hat{\cal W})$ over separable states, is violated by the data set, ultimately proving that entanglement is present among the subsystems. 
The operator $\hat{\cal W}$ is optimal, in that any other normalized linear combination $\hat{\cal W}'=-\sum_a W'_a \hat A_a$ defines an EW inequality whose violation cannot exceed the violation of the inequality involving $\hat{\cal W}$. This property follows from the convexity of the set of separable states. 

\noindent\emph{Complexity of the algorithm.} If the quantum data contain correlation functions involving up to $k$ points, the effective Hamiltonian $H$ contains ${\cal O}(N^k)$ terms; therefore the computational cost of evaluating statistical averages of the kind of Eq.~\eqref{e.Ap} with a precision of $\epsilon$ (using Monte Carlo sampling) scales as ${\cal O}(d^2\epsilon^{-2}N^k)$, where ${\cal O}(d^2)$ is the cost of evaluating the local observables $o_{m_i}^{(i)}(\bm \lambda_i)$ when $d_i=d$. The polynomial scaling of the computational cost with the number $N$ of parties and with the local Hilbert space dimension is the central asset of our approach. 

\noindent \emph{Ensembles of qubits.} Hereafter we shall specify our attention to the case of systems of $N$ qubits partitioned into subsystems consisting of single qubits; and quantum data will be assumed to consist of one- and two-point correlations, $\langle \hat\sigma_a^{(i)} \rangle_{\hat\rho}$ and  $\langle \hat\sigma_a^{(i)} \hat\sigma_b^{(j)} \rangle_{\hat\rho}$ respectively, fully specifying all one- and two-qubit reduced density matrices. Product states are parametrized by the orientations $\{\bm \lambda_i \}= \{\bm n^{(i)} \}$ of each qubit on the Bloch sphere (where $\bm n^{(i)}$ are unit vectors), so that the effective Hamiltonian describes \emph{classical Heisenberg spins} (namely, rotators), coupled via bilinear interactions and immersed in an external field: 
\begin{equation}
	H(\{{\bm n}^{(i)}\}) = -\sum_{i=1}^N \sum_{a=x,y,z} K_a^{(i)} n_a^{(i)} \nonumber \\
	- \sum_{i<j} \sum_{a,b} K_{ab}^{(ij)} n_a^{(i)} n_b^{(j)}~.
	\label{eq_H_cl} 
\end{equation}

\begin{figure}
\includegraphics[width=\linewidth]{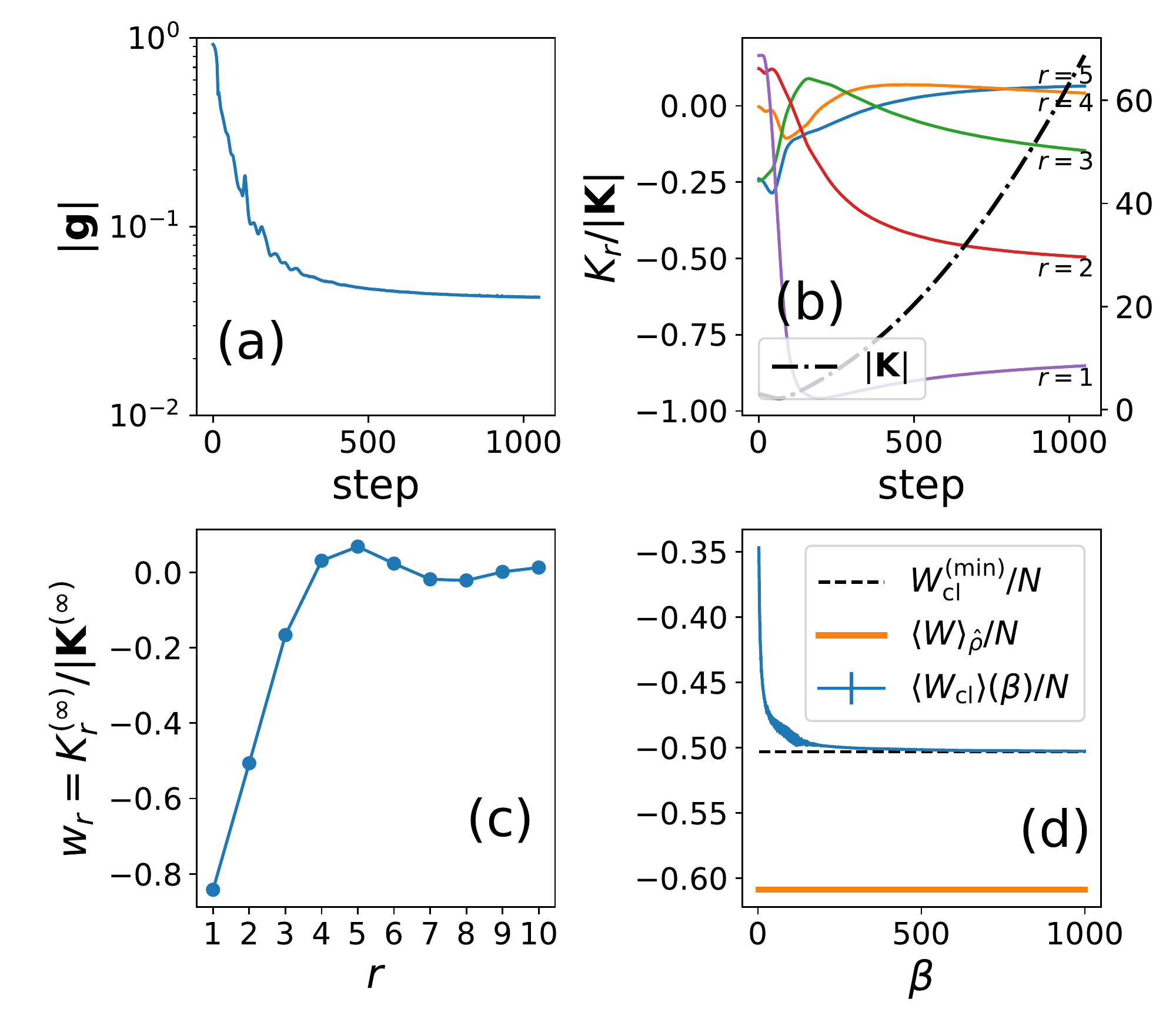}
\caption{Data-driven entanglement witness for the Heisenberg chain at $T/J=1$. (a) Distance between the quantum data (all spin-spin correlators) and the optimized separable state (${\bm g}$: gradient of the cost function), as a function of optimization steps in a Nesterov accelerated gradient descent ($\epsilon=0.01$). Each step contains $10^5 - 10^7$ Monte Carlo steps to achieve a relative precision of $10\%$ on the modulus of the gradient \cite{SM}. (b) Normalized coupling constants $K_r$ in the classical Hamiltonian defining the separable state (solid lines, left axis), and overall amplitude $|{\bm K}|$ (dashed-dotted line, right axis). (c) Normalized couplings $K_r$ at the end of the algorithm; (d) The separable bound can be obtained via simulated annealing \cite{Kirkpatricketal1987} by calculating $\langle {\cal W}_{\rm cl} \rangle(\beta)$ against $\exp[-\beta {\cal W}_{\rm cl}]$, ramping $\beta$ from $0$ to $1000$. The minimum ${\cal W}_{\rm cl}^{({\rm min})}$ is actually the lowest value recorded for ${\cal W}_{\rm cl}$ throughout the ramp.}
\label{fig2}
\end{figure}

\noindent \textit{Heisenberg antiferromagnetic chain.} The first example of entangled states that we study with our approach is the thermal equilibrium state of the $S=1/2$ Heisenberg chain $\hat H = J \sum_{i=1}^N \hat{\bm S}^{(i)} \cdot \hat{\bm S}^{(i+1)}$, where $\hat{\bm S}^{(i)}$ are $S=1/2$ spin operators, $J$ is the exchange energy, and periodic boundary conditions (PBC) are assumed. Thermal equilibrium states $\hat\rho$ ($ \propto \exp[-\hat H / k_BT]$) give $\langle {\bm \hat\sigma}_a^{(i)} \rangle_{\hat\rho} = 0$ and $\langle \hat\sigma_a^{(i)} \hat\sigma_b^{(j)} \rangle_{\hat\rho} = \delta_{ab}~ C(|i-j|)$, due to rotational invariance of the spin-spin couplings and translational invariance. These elementary symmetries of the quantum data are directly inherited by the classical Hamiltonian defining separable states aimed at reproducing them. The Hamiltonian takes the form of a classical long-range Heisenberg model $H(\{{\bm n}^{(i)}\}) = -\sum_{i<j} K_{|i-j|} {\bm n}^{(i)} \cdot {\bm n}^{(j)}$ with $K_r = K_{N-r}$. 
The most effective existing multipartite entanglement criterion for this quantum data is based on the collective spin, namely $\langle \hat{\bm J}^2 \rangle = \sum_{ij} \langle \hat{\bm S}^{(i)}\cdot \hat{\bm S}^{(j)} \rangle < N/2$ \cite{Toth2004,Wiesniaketal2005}, which is verified for  $t = T/J \lesssim 1.4$. This criterion is a permutationally invariant EW (PIEW), treating correlations at all distances on the same footing, and it cannot be optimal at sufficiently high temperatures, namely when the correlation length $\xi$ becomes of the order of a few lattice spacings.    

As a first validation of our approach, we search for the optimal EW based on two-body correlations $\langle \hat\sigma_a^{(i)} \hat\sigma_a^{(j)}\rangle$ by using as input quantum data the correlations (obtained via quantum Monte Carlo - QMC \cite{SM}) at $t=1$ for $N=64$ spins, at which $\xi = 0.72$. Because of their finite range we only used correlations up to a distance $r_{\rm max}=10$.
Fig.~\ref{fig2} illustrates the results of our approach. The saturation to a finite value of the distance between the quantum data and those of the optimized separable state (measured by the norm of the vector ${\bm g}$, see Fig.~\ref{fig2}(a)) and the divergence of the couplings $K_r$ (Fig.~\ref{fig2}(b)) clearly indicate the success of entanglement witnessing. 
The optimal EW operator can be reconstructed in principle from the asymptotic value of the gradient vector ${\bm g}^{(\infty)}$ as $\hat{\cal W} = -\sum_{i=1}^N \sum_{a \in \{x,y,z\}} \sum_{r=1}^{r_{\max}} w_r ~ \hat\sigma_a^{(i)} \hat\sigma_a^{(i+r)}$ with $w_r = -g_r^{(\infty)} / |{\bm g}^{(\infty)}|$. In practice, we found a more strongly violated EW inequality using the asymptotic couplings of the effective Hamiltonian, namely $w_r = K_r^{(\infty)} / |{\bm K}^{(\infty)}|$ -- which display a clear spatial structure, shown in Fig.~\ref{fig2}(c) (see \cite{SM} for the numerical values). 
The final step of the approach consists in determining the separable bound $B_{\rm sep} = \min_{\hat\rho_p}{\rm Tr}(\hat\rho_p \hat{\cal W})$. The latter can be obtained as the solution of a set of algebraic equations \cite{sperlingetal2013,gerkeetal2018}; here we rather obtain it by finding the ground-state energy of the classical Hamiltonian ${\cal W}_{\rm cl} = -\sum_{i=1}^N \sum_{r=1}^{r_{\max}} w_r ~ {\bm n}^{(i)} \cdot {\bm n}^{(i+r)}$ via temperature annealing \cite{Kirkpatricketal1987} [Fig.~\ref{fig2}(d)]. We observe that $B_{\rm sep}/N = -0.5032$, while the quantum data reach $\langle \hat{\cal W}\rangle_{\hat\rho}/N = -0.6089$. In contrast, the best PIEW -- properly normalized \cite{SM} -- is violated by an amount of $0.04552$. This result is \emph{not} incremental, because the EW inequality we find is optimal among all those containing two-body correlators. Interestingly, for temperatures $t\gtrsim 1.4$ (at which the PIEW ceases to work) we found numerically impossible to prove that $\hat\rho(T)$ is entangled solely based on two-point correlators: this in turn shows that the maximal set of thermal states whose entanglement can be witnessed using two-point correlators is essentially captured by the PIEW. This will not be the case in our next example, in which our approach significantly extends the range of witnessed entangled thermal states.

\begin{figure}
\includegraphics[width=1\linewidth]{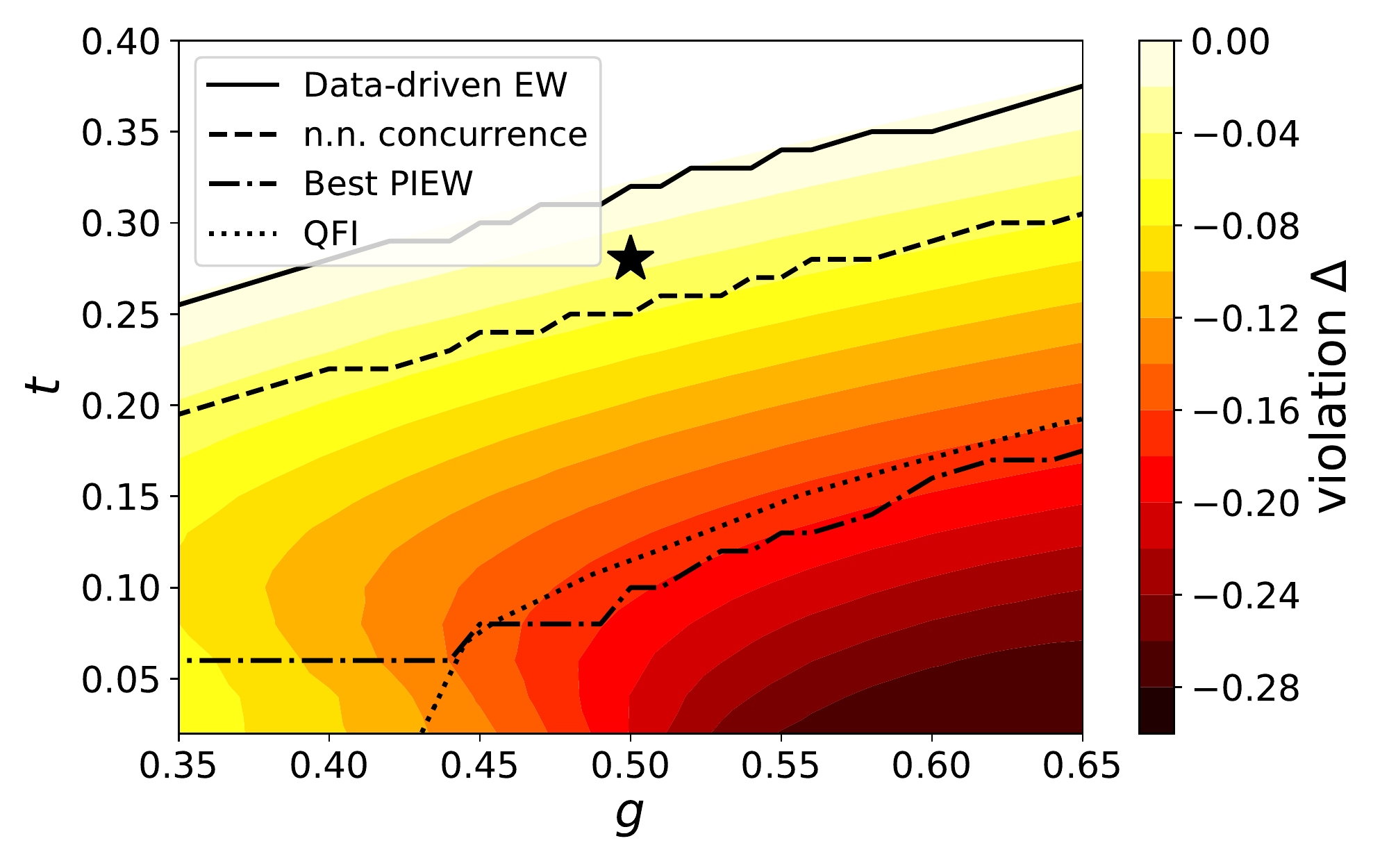}
\caption{Data-driven EW for the quantum Ising chain. Phase diagram around the QCP. The star corresponds to $t=0.28$, $g=0.5$, at which the quantum data used as input were calculated. The color represents the violation $\Delta = (\langle \hat{\cal W} \rangle_{\hat\rho} - B_{\rm sep})/N$ of our data-driven EW. The various curves correspond to the temperature below which different entanglement criteria are satisfied (nearest-neighbour concurrence \cite{wooters1998}; best PIEW \cite{tothetal2009}; and quantum Fisher information (QFI) of $\hat J_z$ \cite{haukeetal2016}.}
\label{fig_EW_Ising}
\end{figure}
\noindent \textit{Quantum Ising chain.} Our final example is the quantum Ising model with Hamiltonian $\hat H = -J \sum_{i=1}^N (\hat S_z^{(i)} \hat S_z^{(i+1)} + g \hat S_x^{(i)})$, where $J$ is the interaction strength and $Jg$ the transverse field. In the ground state, the system displays a quantum critical point (QCP) at $g=g_c=1/2$ between a ferromagnetic phase ($g<g_c$) and a paramagnetic phase ($g>g_c$) \cite{sachdev2011quantum}. At finite temperature around the QCP, the system is known to exhibit robust entanglement \cite{haukeetal2016,PhysRevLett.121.020402,frerotR2018}.  Given the symmetries of the correlation functions ($\langle \hat\sigma^{(i)}_a \rangle_{\hat\rho}=0$ for $a = y,z$; $\langle \hat\sigma_a^{(i)} \hat\sigma_b^{(j)} \rangle_{\hat\rho} \sim \delta_{ab}$), the classical Hamiltonian tailored to reproduce them is of the form: $H(\{{\bm n}^{(i)}\}) = -K_x \sum_{i=1}^N n_x^{(i)} -\sum_{a=x,y,z}\sum_{i<j} K_{a}^{|i-j|} n_a^{(i)} n_a^{(j)}$. As input quantum data, we consider the correlation functions of a chain of $N=64$ spins with PBC at a temperature $t=T/J=0.28$ for $g=0.5$ - obtained as well via QMC. Given the finite correlation length, we only used correlators up to a distance $r_{\rm max}=20$. Following the same procedure as described for the Heisenberg chain, we find an optimal EW operator which is spatially structured, of the form $\hat {\cal W}= -w_x \sum_{i=1}^N \hat\sigma_x^{(i)} -\sum_{a=x,y,z}\sum_{i<j} w_{a}^{(|i-j|)} \hat\sigma_a^{(i)} \hat\sigma_a^{(j)}$ (coefficients and separable bound in the Supplemental Material \cite{SM}). On Fig.~\ref{fig_EW_Ising}, we show that this new EW criterion, optimal for the thermal state at $t=0.28,g=0.5$, allows one to prove entanglement for a larger set of thermal states than all the existing criteria in the literature (namely the nearest-neighbour concurrence \cite{wooters1998}, the PIEW \cite{tothetal2009}, and the quantum Fisher information \cite{haukeetal2016} -- see \cite{SM} for further details).

\noindent \textit{Conclusions.}
We introduced a data-driven method to probe multipartite entanglement in many-body systems. This method relies on mapping separable states onto Boltzmann distributions for a classical field theory on a lattice. The classical degrees of freedom of this field theory are dictated by the considered partitioning of the system. The structure of the corresponding classical Hamiltonian is dictated by the quantum data at hand; and its parameters are optimized in order to fit at best the quantum data. This method allows to exhaustively test the entanglement witnessing capability of a set of quantum data in a time scaling polynomially with the number of parties in the partition (if the size of quantum data is also polynomial); this is guaranteed whenever the classical field theory is not a model of a glass (namely when it does not feature disorder \emph{and} frustration).  This opens the way to the systematic certification of entanglement in intermediate-scale quantum devices.

\acknowledgments{We are very grateful to Antonio Ac{\'i}n for insightful discussions. IF acknowledges support from the Government of Spain (FIS2020-TRANQI and Severo Ochoa CEX2019-000910-S), Fundaci{\'o} Cellex and Fundaci{\'o} Mir-Puig through an ICFO-MPQ Postdoctoral Fellowship, Generalitat de Catalunya (CERCA, AGAUR SGR 1381 and QuantumCAT), and MINECO-EU QuantERA MAQS (funded by The State Research Agency (AEI) PCI2019-111828-2 / 10.13039/501100011033). TR acknowledges support from ANR (EELS project) and QuantERA (MAQS project). Numerical computations have been performed at the P{\^o}le Scientifique de Mod{\'e}lisation Num{\'e}rique (PSMN).

\bibliography{biblio}

\begin{thebibliography}{10}
\bibitem{footnote1}
\bibinfo{title}{Otherwise one can always reproduce the correlations among these
  observables using separable states by choosing the local states
  $|\psi_i\rangle$ to be joint eigenstates of the observables, and reproducing
  with $p$ the classical correlations among the observables.}
\end{thebibliography}

\appendix
\begin{center}
\large{\textbf{Supplemental Material}}
\end{center}

In this Supplemental Material, we provide: 1) further technical details on the variational algorithm described and implemented for the data presented in the main text; 2) on the generation of quantum data, used as input to our algorithm, by quantum Monte Carlo; 3) on the relative versus absolute violation of the entanglement witnesses; 4) on the comparison with existing entanglement criteria. In the attached \texttt{.csv} files, the numerical coefficients of the entanglement witnesses discussed in the main text are given.

\section{Details on the algorithm}
\subsection{Mapping of the separability problem onto an inverse statistical problem}

\emph{General strategy.}
Following the notations of the main text, we assume that the quantum data consist of a collection of expectation values $\{\langle \hat A \rangle_\rho\}_{a=1}^R$ measured on the quantum device. Our constructive strategy to solve the separability problem is to try and reproduce these data with a separable state -- the failure of this   attempt marking the success of entanglement detection. As discussed in the main text, a separable state $\hat\rho_p$ is defined by an arbitrary probability distribution $p(\bm\lambda)$ over local quantum states $|\psi_i(\bm \lambda_i)\rangle$. Our strategy is then to build an optimal $p_{\rm opt}(\bm\lambda)$, such that the corresponding separable state $\hat\rho_{p_{\rm opt}}$ produces the best possible approximation to the available data attainable using separable states. For a given separable state $\hat\rho_p$, we have:
 \begin{equation}
\langle \hat A_a\rangle_{\hat\rho_p} = \int d\bm\lambda ~p(\bm \lambda) {\cal A}_a(\bm \lambda) =: \langle {\cal A}_a \rangle_p
 \end{equation}
 where ${\cal A}_a(\bm \lambda) = \sum_{\bm m} x_{\bm m}^{(a)} \prod_{i=1}^N o_{m_i}^{(i)}(\bm \lambda_i)$ and $o_{m_i}^{(i)}(\bm \lambda_i) = \langle \psi_i(\bm \lambda_i) | \hat O_{m_i}^{(i)} | \psi_i(\bm \lambda_i) \rangle$ (see the main text for the precise definition of the $\hat O$'s operators). 
 
 Our approach is in essence a \textit{variational approach}, in which we parametrize the distribution $p(\bm \lambda)$ as a Boltzmann distribution $p(\bm\lambda) = \exp[-H(\bm\lambda)] / {\cal Z}({\bm K})$ associated with a classical Hamiltonian $H(\bm \lambda) =-\sum_{a=1}^R K_a {\cal A}_a(\bm \lambda)$. Two crucial properties, on which we further elaborate in this section, make this choice of Ansatz especially suited to solve the separability problem. Firstly, the expressive power of this Ansatz is \textit{complete}, which means that there is no loss of generality in looking for a separable state of this specific form: if a separable state of this form cannot reproduce the data, then no separable state whatsoever can do so. Secondly, the variational parameters $K_a$ can be optimized by minimizing a \textit{convex} cost function, whose gradient can be evaluated at a cost scaling polynomially with $N$ (the number of local subsystems) and with $d$ (the local Hilbert space dimension), under mild assumptions (specifically, the absence of glassiness of the classical model $H(\bm\lambda)$). 
 
\textit{Completeness of the Ansatz.} If a distribution $p(\bm \lambda)$ exists which reproduces the data set: $\langle {\cal A}_a \rangle_p = \langle \hat A \rangle_{\hat\rho}$ for all $a=1, \dots R$, it is generically not unique. One may therefore choose the distribution which, as a further specification, maximizes the Shannon entropy functional $S[p] = -\int d{\bm \lambda} ~p(\bm\lambda) \log p(\bm\lambda)$. This amounts to removing any other constraints on the distribution except that of reproducing the data set with its averages.
 Following the seminal work of Jaynes \cite{jaynes1957}, maximizing $S[p]$ under the constraint of reproducing the data is achieved upon introducing Lagrange multipliers $K_a$, and minimizing the functional $F[p] = -S[p] - \sum_{a=1}^R K_a [\langle {\cal A}_a \rangle_p - \langle \hat A_a \rangle_{\hat\rho}]$. Setting to zero the functional derivative with respect to $p(\bm\lambda)$ yields as a solution the Boltzmann distribution $p(\bm\lambda) = \exp[\sum_a K_a {\cal A}_a(\bm\lambda)] / {\cal Z}({\bm K})$.  The parameters $K_a$ are hence exactly the tuning knobs that allow $p(\bm \lambda)$ to satisfy the constraints to the best that a classical probability distribution can do.
 
 To further understand this point, let us stress that throughout our work we assume (as it is reasonable to do) that the size of the quantum data set scales at most polynomially with system size, so that the number of constraints associated with the reproduction of the quantum data also scales polynomially.
On the other hand a distribution $p(\bm \lambda)$ is uniquely defined by an exponentially large number of constrains -- as many as the possible values of the argument $\bm \lambda$. 
The exponentially many constraints, to be added in order to specify the distribution uniquely, cannot help it in any way in reproducing the quantum data. On the other hand, maximizing the entropy of the distribution precisely gets rid of the useless constraints beyond the ones imposed by the quantum data themselves.
Once the least constrained distribution is achieved upon maximizing the entropy (subject to the constraint), varying the parameters $K_a$ of the distribution exactly allows one to reproduce all the data sets which could potentially be produced by the most general distribution $p(\bm\lambda)$. The Boltzmann distribution associated with the classical Hamiltonian $H(\bm\lambda)=-\sum_a K_a {\cal A}_a(\bm\lambda)$ can therefore be viewed as an Ansatz whose expressive power of quantum data sets of is as high as one can possibly achieve with a classical distribution.

\textit{Optimizing the variational parameters.} We then show that the parameters $K_a$ can be optimized upon minimizing a \textit{convex} cost function. Convexity is a crucial property of the whole procedure: if the optimization relied on a heuristic algorithm, then the failure to reproduce the quantum data could simply mean that the optimization has been trapped in some local minimum \cite{boyd2004convex}, and therefore the result would be inconclusive. As stated in the main text, a convex cost function for our problem is given by $L({\bm K}) = \log {\cal Z}({\bm K}) - \sum_a K_a \langle \hat A_a \rangle_{\hat\rho}$. Another crucial aspect for the scalability of our algorithm is that the cost function $L({\bm K})$ itself is never computed. Only its gradient $g_a=\partial L/\partial K_a = \langle {\cal A}_a \rangle_p - \langle \hat A_a \rangle_{\hat \rho}$ is evaluated, and used to update the parameters $K_a$ in a gradient-descent algorithm, or any improvement thereof (in this paper, we used the accelerated gradient-descent algorithm of Nesterov). Even though the cost function itself is never computed, its existence and its convexity are key to ensure the monotonous convergence of our algorithm towards the global optimum of the problem \cite{boyd2004convex}. Specifically, together with the cost function, the norm of its gradient converges towards its minimal value; namely, the distribution $p(\bm\lambda)$ converges towards the best possible approximation to the data with a separable state. If a distribution $p(\bm\lambda)$ exists which reproduces the data, then the gradient of the cost function vanishes, and it is impossible to detect entanglement from the available data. Notice that this is not a limitation of our approach, but on the contrary it represents a fundamental property of the data that our method exhibits. On the other hand, if the data lie outside of the convex region reachable by separable states, the cost function $L$ is not bounded from below, and the gradient will stabilize to a finite value, leading to a runaway to infinity of the coupling constants $K_a$, and marking the success of entanglement detection -- as further discussed in the main text. 

\textit{Computational complexity.} Finally, we would like to remark that the computational cost required to evaluate the gradient $g_a = \langle {\cal A}_a \rangle_p - \langle \hat A_a \rangle_{\hat \rho}$ with a given relative precision of $\epsilon$ via Monte Carlo methods scales as $1/\epsilon^2$. One could imagine a fine-tuned situation in which the distance between the data under investigation and the separable set is exponentially small in the system size: $|{\bm g}|={\cal O}[\exp(-N)]$, which would translate into a computational cost of our algorithm diverging exponentially with $N$. While such a situation cannot be excluded \textit{a priori}, in any practical instance the quantum data come with a finite uncertainty -- certainly not decreasing exponentially with the system size. Indeed, the best scaling of the relative uncertainty that one can expect is as $N^{-1/2}$, when considering collective observables which are the sums of ${\cal O}(N)$ nearly independent degrees of freedom (as it happens in systems with a finite correlation length), and the same benign scaling is shared by Monte Carlo estimates of the same quantities. On the other hand, exponentially decreasing precision would require exponentially large statistics, which is not a realistic assumption for any source of the quantum data set (be it experiments or numerical calculations).
As a consequence, quantum data whose distance to the separable set scales exponentially with system size would inevitably be reproduced by our algorithm using a separable state \emph{within their uncertainty}, and at a polynomial cost.

In the literature, the separability problem has been proved to be NP-hard in the bipartite case \cite{gurvits2003}. This implies that there exists instances requiring an exponential cost in the local Hilbert space dimension $d$. On the other hand, we are not aware of a similar complexity result in the multipartite case, namely for a fixed $d$ ($d=2$ in the qubit examples treated explicitly in this work), but increasing the number $N$ of parties. 
For the multipartite separability problem with $N$ qubits, we state in the main text that classical glassy models define the practical limitation to the scalability of our approach. We would like to emphasize that this assumption is rather conservative. Indeed, the classical models one has to sample in our approach involve continuous degrees of freedom (e.g. $N$ classical rotators representing vectors on the Bloch sphere, defining the local quantum states, see Section \ref{sec_multipartite_qubits} below). While Ising spin glasses, which involve $\pm 1$ variables, have been proved to be NP-hard \cite{Barahona1982}, a similar result does not exist for frustrated disordered classical models involving rotators (to the best of our knowledge). This (classical) statistical-physics observation is consistent with the absence of formal proof of NP-hardness of the (quantum) multipartite separability problem.

Concerning the bipartite case ($N=2$, increasing $d$), whose NP-hardness is proven \cite{gurvits2003}, our algorithm has a cost which is polynomial in $d$, in apparent contradiction with the complexity result. First, we notice that the NP-hardness \cite{gurvits2003} concerns the situation where the full bipartite state $\rho_{AB}$ (which is a $d^2 \times d^2$ Hermitian matrix of unit trace) is used as input. Our algorithm treats a more general situation, where 1- and 2-body correlations of the form $\langle \hat A_a \rangle$, $\langle \hat B_b \rangle$, $\langle \hat A_a \hat B_b \rangle$ are known [where $\hat A_a$ ($a\in \{1, \dots, R_A\}$), and $\hat A_b$ ($a\in \{1, \dots, R_B\}$) are local observables on $A$ and $B$ subsystems, respectively]. This knowledge is equivalent to the knowledge of $\rho_{AB}$ if $\hat A_a$ and $\hat B_b$ form tomographically complete sets of observables (for instance, the $R_A=R_B=d^2 - 1$ generalized Gell-Mann matrices, which are the generators of $SU(d)$). In our approach, we parametrize separable states as Boltzmann distributions related to $H(\psi_A, \psi_B) = -\sum_a K_a {\cal A}_a(\psi_A) - \sum_b K_b {\cal B}_b(\psi_B) - \sum_{ab} K_{ab} {\cal A}(\psi_A) {\cal B}_b(\psi_B)$, where $\psi_A$ and $\psi_B$ represent local quantum states, parametrized by $2d-2$ classical variables, and where $K_a, K_b, K_{ab}$ are $R_A + R_B + R_A R_B$ variational parameters. The NP-hardness result \cite{gurvits2003} implies that if one considers tomographically-complete sets of observables, then there exists instances of parameters $K_a, K_b, K_{ab}$ for which sampling the corresponding Boltzmann distribution takes a time diverging exponentially with $d$. We cannot immediately identify to which hard statistical physics problem this situation would correspond -- but certainly such hard instances must exist, as imposed by the result of Ref.~\cite{gurvits2003}. In analogy with glassy problems, for these instances the energy landscape described by $H(\psi_A, \psi_B)$ should display a myriad of local minima separated by energy barriers whose height is proportional to $d$, making the sampling of the model via Monte-Carlo methods inefficient. 

On a more constructive tone, we would like to remark that such complexity results only refer to worst-case instances. In the context of our approach, such worst-case instances could correspond to glassy models, and in the case of translationally invariant data considered in this paper, such glassiness is avoided by construction. Such instances are not expected to be generically encountered when analyzing data from present-day quantum simulators of non-disordered systems. Finally, we would like to emphasize that there is no risk of erroneously concluding that entanglement is present if such hard instances manifest themselves. We have already argued above that realistic quantum data cannot reveal entanglement in the case of exponentially small violations of witness inequalities. In the presence of glassiness, instead, one would be unable to run the simulation forward due to very large error bars in the Monte Carlo evaluation of the expectation values for separable states. As a consequence, one would conclude that entanglement cannot be detected within the accuracy of the method.

\subsection{Special case: partitioning the system into $N$ qubits}
\label{sec_multipartite_qubits}
In this work we introduce a variational algorithm to fit a given data set of expectation values by using separable states. In the case of qubits taken as individual subsystems, separable states are represented without loss of generality as Boltzmann distributions over classical Heisenberg spins ${\bm n}^{(i)}$ on the unit sphere (which represent pure states on the Bloch sphere for individual qubits).  In the examples discussed in the main text, the data set contains one-qubit expectation values $\langle \hat\sigma_a^{(i)} \rangle_{\hat\rho}$ and two-qubit correlations $\langle \hat\sigma_a^{(i)} \hat\sigma_b^{(j)} \rangle_{\hat\rho}$. In the examples we considered (namely, the one-dimensional antiferromagnetic Heisenberg model, and the Ising model in a transverse field, both with periodic boundary conditions), correlations $\langle \hat\sigma_a^{(i)} \hat\sigma_b^{(j)} \rangle_{\hat\rho}$ vanish if $a \neq b$. Since we used translationally invariant chains (with periodic boundary conditions), the one-qubit data reduces to the average magnetization $\langle m_a \rangle_{\hat\rho} = \sum_{i=1}^N\langle \hat\sigma_a^{(i)} \rangle_{\hat\rho} /N$, and the the two-qubit correlations depend only on the inter-qubit distance: $\langle C_a^{(r)} \rangle_{\hat\rho} =  \langle \hat\sigma_a^{(i)} \hat\sigma_a^{(i+r)} \rangle_{\hat\rho}$. In the case of the Heisenberg model, which displays $SU(2)$ invariance, we have $m_a=0$. In this case, we considered as quantum data $\langle C^{(r)} \rangle_{\hat\rho} = \langle C_x^{(r)} + C_y^{(r)} + C_z^{(r)} \rangle_{\hat\rho}$. 

Correspondingly, the classical Hamiltonian aiming at reproducing the quantum data contains one- and two-body interactions terms (the latter truncated beyond a given distance $r_{\rm max}$). For the Heisenberg model, we get 
\begin{equation}
H =  -  \sum_{i} \sum_{r=1}^{r_{\rm  max}} K^{(r)}  \bm n^{(i)} \cdot \bm n^{(i+r)}~;
\end{equation}
 while for the 	quantum Ising model, where $m_y=m_z=0$, we have 
 \begin{equation}
 H = - K_x \sum_{i=1}^N n_x^{(i)} - \sum_{a=x,y,z}  \sum_{i} \sum_{r=1}^{r_{\rm  max}}   K_a^{(r)} n_a^{(i)} n_a^{(i+r)}~.
 \end{equation} 
  The $K$'s coefficients are the variational parameters of our algorithm, which are optimized in an iterative manner. A simple gradient-descent algorithm consists in iterating the following update rule (for the Ising model):
\begin{eqnarray}
	K'_{x} & = & K_{x} - \epsilon [\langle m_x \rangle_{p} - \langle m_x \rangle_{\hat\rho}] \\
	(K_{a}^{(r)})' &  =  & K_{a}^{(r)} - \epsilon [\langle C_a^{(r)} \rangle_{p} - \langle C_a^{(r)} \rangle_{\hat\rho}] ~
\end{eqnarray}
for $a\in \{x,y,z\}$, and $r \in \{1, 2, \cdots N/2\}$; 
and (for the Heisenberg model):
\begin{equation}
	(K^{(r)})' = K^{(r)} - \epsilon [\langle C^{(r)} \rangle_{p} - \langle C^{(r)} \rangle_{\hat\rho}] ~.
\end{equation}
In the above equations, $\langle \cdot \rangle_{p}$ is the expectation value on the Boltzmann distribution for the classical Hamiltonian (whose couplings are the $K$'s coefficients), while $\langle \cdot \rangle_{\hat\rho}$ are the target quantum data. As discussed in the main text (see also \cite{frerotR2020}), $\epsilon$ is a small parameter, implementing a numerical gradient descent of the (convex) $L$ function. In practice, we implemented the Nesterov's accelerated gradient-descent (NAG) algorithm, with $\epsilon = 0.01$. 

Each step of the NAG algorithm requires to compute the Euclidean distance  ${\bm g}$ between the separable data and the quantum data, namely to compute $\langle m_x \rangle_{p}$ and $\langle C_a^{(r)} \rangle_{p}$ for the Ising model and $\langle C^{(r)} \rangle_{p}$ for the Heisenberg model. This was implemented using Markov-chain Monte Carlo. The number of Monte Carlo steps (defined below) implemented at each step of the NAG algorithm was chosen such that the relative error on ${\bm g}$ be smaller than a given threshold $\eta$, which we chose as $\eta = 0.05$ for the Ising model, and $\eta = 0.1$ for the Heisenberg model. In other words, one step of the NAG algorithm is completed when:
\begin{equation}
	\frac{2\sum_\alpha |g_\alpha|~ {\rm Err}(g_\alpha)}{|{\bm g}|^2} < \eta ~,
\end{equation}
where ${\rm Err}(g_\alpha)$ is the error on $g_\alpha$, as estimated from the Monte Carlo algorithm. 
 Each step of the Monte Carlo algorihm consisted of $2N$ iterations of single-spin Metropolis updates and of single-spin microcanonical overrelaxation updates \cite{Creutz1987}.  
 The amplitude of the proposed Metropolis updates was adapted along the Monte Carlo simulation so that the move be accepted with frequency $0.5 \pm 0.1$.
 Therefore, a single Monte Carlo step consists of $2N$ microcanonical updates, and of  $N$ accepted Metropolis updates (on average). 
 
 As the variational optimization of the $K$'s parameters progresses along the NAG algorithm, the norm of the gradient ${\bm g}$ decreases, and therefore an increasing number of Monte Carlo steps is required at each step of the NAG algorithm in order to achieve the required relative precision of $\eta$. When the quantum data cannot be fitted by a separable state, ${\bm g}$ stabilizes to a finite value. The number of steps of the NAG algorithm to achieve this convergence (and therefore the total number of Monte Carlo steps along the whole optimization) depends on the value of $|{\bm g}|$ as obtained at the end of the optimization. For the examples presented in the main text, about $10^3$ steps of the NAG algorithm were necessary, each of them comprising  $10^4 \div 10^7$ Monte Carlo steps. 

\section{Quantum data from Quantum Monte Carlo}

Data-driven entanglement witnessing is fundamentally based on reliable quantum data on quantum many-body systems. Here we chose to use quantum Monte Carlo data for quantum spin chains at finite temperature, obtained using Stochastic Series Expansion \cite{SyljuasenS2002}, which provides numerically exact correlation functions for the model of interest (within the statistical error bar). Finite-temperature equilibrium calculations offer the most reliable source of data for mixed states -- which pose the real challenge in terms of entanglement detection, while for pure states any form of connected correlation is an entanglement witness.  Beyond their significance in condensed matter physics and quantum statistical physics, the models we chose (quantum Heisenberg and quantum Ising chain) are also of direct relevance to several experiments in quantum simulation, see e.g. \cite{Bolletal2016, BrowaeysL2020} for recent examples.  

\section{Existing entanglement witnesses}
In this section, we provide additional details on the existing entanglement witnesses against which the quantum data of the quantum Ising model were tested (Fig.~3 of the main text). \\

\noindent \textit{Concurrence.} The concurrence \cite{wooters1998} defines a necessary and sufficient condition for the separability of a two-qubits density matrix. We computed the concurrence between nearest-neighbours, after reconstructing the density matrix $\hat\rho_{12}$ from the knowledge of one- and two-qubits expectation values $\langle \hat\sigma_a^{(1)} \rangle_{\hat\rho}$, $\langle \hat\sigma_b^{(2)} \rangle_{\hat\rho}$ and $\langle \hat\sigma_a^{(1)} \hat\sigma_b^{(2)} \rangle_{\hat\rho}$ (with $a,b \in \{x,y,z\}$) \cite{paris2004quantum}. The dashed line on Fig.~3 defines the temperature below which $\hat\rho_{12}$ is entangled. Since the concurrence criterion \cite{wooters1998} is based on a subset of the full quantum data we considered (which contains all one- and two-qubits correlations functions, which is equivalent to all two-body reduced density matrices $\hat\rho_{ij}$, and not only $\hat\rho_{12}$), by construction our data-driven method must detect entanglement in a region of the phase diagram strictly larger than the one detected by the concurrence -- a fact clearly visible on Fig.~3.  \\

\noindent \textit{Permutationally-invariant entanglement witnesses.} In Ref.~\cite{tothetal2009}, a complete family of 8 entanglement witnesses based on the two-qubits reduced density matrix averaged over all pairs, $\hat\rho_{\rm av,2} = 2\sum_{i \neq j} \hat\rho_{ij} / [N(N-1)]$, was derived. Equivalently, $\hat\rho_{\rm av,2}$ is reconstructed from the knowledge of all one- and two-body correlations averaged over all permutations: $m_a := \sum_{i=1}^N \langle \hat\sigma_a^{(i)} \rangle_{\hat\rho}$ and $C_{ab} := \sum_{i \neq j} \langle \hat\sigma_a^{(i)} \hat\sigma_b^{(i)} \rangle_{\hat\rho}$. Since $m_a$ and $C_{ab}$ are coarse-grained features of the quantum data we have considered, if an EW inequality is violated by $m_a$ and $C_{ab}$ (namely if one of the 8 EW inequalities of ref.~\cite{tothetal2009} is violated), then our data-driven algorithm must also reconstruct a violated entanglement witnesses -- in general, a more strongly violated one. As illustrated on Fig.~3 for the quantum Ising model, for which we tested all 8 criteria for each parameters $(t,g)$ (temperature and transverse field), this is clearly the case.  \\

\noindent \textit{Quantum Fisher information.} The quantum Fisher information (QFI) is another multipartite entanglement witness. Formally, the QFI quantifies the sensitivity of the state $\rho$ to unitary transformations $\hat\rho(\phi) = e^{-i\phi \hat O} \hat\rho e^{i\phi \hat O}$ with $\hat O$ a quantum observable \cite{pezzeS2014}. The QFI can be expressed as ${\rm QFI}(\hat O, \hat \rho) =2 \sum_{n \neq m} (p_n - p_m)^2 |\langle n | \hat O |m \rangle|^2 / (p_n + p_m)$, where $\hat \rho$ is diagonalized as $\hat \rho = \sum_n p_n |n\rangle \langle n |$.  Here, we chose for $\hat O$ the collective spin $J_z = \sum_{i=1}^N \hat\sigma_z^{(i)} / 2$, which is optimal to witness entanglement around the quantum critical point of the quantum Ising model \cite{haukeetal2016,PhysRevLett.121.020402}. The inequality ${\rm QFI}(\hat J_z, \hat \rho) \leq N$  is satisfied by all separable states, so that a QFI exceeding the system size is an entanglement witness \cite{pezzeS2014}. In general, computing the QFI involves the knowledge of the full density matrix $\rho$, but the mapping of the quantum Ising chain onto a free-fermion model \cite{sachdev2011quantum} makes this computation tractable \cite{haukeetal2016}. Notice that computing the QFI requires knowledge beyond one- and two-body correlators, and therefore it goes beyond the data set we have considered. Hence there is no guarantee a priori that our method exceeds the EW capability of the QFI. 
Nevertheless, as illustrated on Fig.~3, the parameter region where entanglement is detected by the QFI is broadly included in the region where entanglement is detected via our data-driven algorithm. 

\section{Absolute versus relative violation of the entanglement witnesses}
By construction, the optimal witness found by our approach is the one whose absolute violation is maximized. Namely, among all possible witness operators $\hat {\cal W} = -\sum_{a} W_a \hat A_a$, properly normalized with the euclidian norm $\sum_a W_a^2=1$, our witness operator maximizes the \textit{difference} $B_{\rm sep} - {\rm Tr}(\hat \rho \hat {\cal W})$, where $B_{\rm sep} = \min_{\hat \rho_{\rm sep}} {\rm Tr}(\hat \rho_{\rm sep} \hat {\cal W})$. As a consequence, it is the witness operator which is most robust to the uncertainty present on the quantum data, namely the one that requires the lowest amount of statistics producing the quantum data themselves.

On the other hand, it is also relevant to consider the \textit{noise robustness} of a given entanglement witness, namely the robustness to a noisy, imperfect preparation of the quantum state $\hat\rho$ that should produce the quantum data. Noisy state preparation can be generically modeled as turning the target state into $(1-\eta) \hat \rho + \eta \mathbb{1}/D$, where $\eta$ parametrizes the strength of the noise, and $D$ is the total Hilber space dimension. Assuming that all operators $\hat A_a$ composing the witness are traceless (which is the case for tensor products of local Pauli matrices, as considered in this paper), this leads us to define the noise robustness as the maximal value of $\eta$ such that $(1-\eta_{\rm max}) {\rm Tr}(\hat \rho \hat {\cal W}) = B_{\rm sep}$, namely $\eta_{\rm max} = 1 - B_{\rm sep} / {\rm Tr}(\hat \rho \hat {\cal W})$. There is no guarantee that the witnesses found by our approach are those whose noise robustness is maximal, and in fact it turns out not to be the case, as shown by the following example.

In the case of the Heisenberg chain, we have considered translationally-invariant entanglement witnesses of the form $\hat {\cal W} = -\sum_{a \in \{x,y,z\}} \sum_{i=1}^N \sum_{r=1}^{r_{\rm max}} w_r \hat \sigma^{(i)}_a \hat \sigma^{(i+r)}_a$. Our convention has been to normalize them to $\sum_r w_r^2 = 1$. For a meaningful comparison with the PIEW $ \langle[\sum_{i=1}^N \hat{\bm S}^{(i)}]^2\rangle \ge N/2$, the latter should be properly normalized according to the same convention, namely: $\hat {\cal W}_{\rm PIEW} = (N-1)^{-1/2} \sum_{a \in \{x,y,z\}} \sum_{i=1}^N \sum_{j \neq i} \hat \sigma^{(i)}_a \hat \sigma^{(j)}_a$, with a separable bound given by $-N  / \sqrt{N-1}$. For the data considered in the main text (namely, a thermal state of the one-dimensional Heisenberg model at temperature $T/J=1.00$ with $N=64$ spins), we find a violation $-N/\sqrt{N-1} - {\rm Tr}(\hat \rho \hat {\cal W}_{\rm PIEW}) = 0.04552$. In contrast, the optimal witness found by our data-driven approach exhibits a larger violation of $0.10570$. On the other hand, the noise robustness of the PIEW is $\eta_{\rm max} = 0.255$, while the noise robustness of the data-driven EW found by our approach is $\eta_{\rm max} = 0.174$. This is qualitatively consistent with the observation that the PIEW is violated up a temperature of $T/J \approx 1.400$, which is higher than the temperature up to which the data-driven EW (optimal by construction at $T/J = 1.00$) is violated.
\\

\section{Detailed numerical values of the entanglement witnesses}
The numerical coefficients of the entanglement witnesses reconstructed by our algorithm are given in this Section. For the Heisenberg model at temperature $T/J=1$ (Fig.~2 of the main text), we discarded the correlations at distances beyond $r=10$. The coefficients $K^{(r)}$ of the entanglement witness are given in Table \ref{tab_dd_heisenberg}. The separable bound is $E/N = -0.503248446$ ($N=64$). 
\begin{table}[h]
\begin{tabular}{c|c}
 distance & $K^{(r)}$ \\
 \hline 
 1 & -0.84192229 \\
 2 & -0.50632705 \\
 3 & -0.16643027 \\
 4 &  0.03072345 \\
 5 &  0.06820392 \\
 6 &  0.02322838 \\
 7 & -0.01850183 \\
 8 & -0.02146924 \\
 9 &  0.00126596 \\
 10 &  0.01267421
\end{tabular}
\caption{Coefficients of the data-driven entanglement witness presented on Fig.~2 of the main text (Heisenberg chain $T/J=1.00$).}
\label{tab_dd_heisenberg}
\end{table}
For the quantum Ising model (Fig.~3 of the main text), the corresponding coefficients $K_x$, $K_x^{(r)}$, $K_y^{(r)}$ and $K_z^{(r)}$ are presented on Table \ref{tab_dd_Ising}. The separable bound is $E/N =-0.465475151529285$ ($N=64$). 

\begin{table*}
\begin{tabular}{c|c|c|c|c}
distance&$K_x$&$K_{xx}$&$K_{yy}$&$K_{zz}$ \\
0&0.078863109939705&&&\\
1&&0.336212178177562&-0.53946940308646&0.701821871161535\\
2&&0.006136971013906&-0.186782505835586&-0.213519552791139\\
3&&-0.001090090530485&-0.00257250626173&-0.098573694386247\\
4&&-0.016994651046548&0.023567559817685&0.057667124559048\\
5&&0.003961220213939&-0.003352864510561&0.032995580923791\\
6&&0.006728876469642&-0.006668943216078&-0.024548154286714\\
7&&-0.001277958269768&0.003102691532182&-0.016751237825024\\
8&&-0.000351393929815&0.001659840517445&0.012307823520514\\
9&&0.002460403552907&-0.001929489352437&0.010912062898113\\
10&&0.000818810042566&5.60569829502012E-05&-0.007413177478221\\
11&&0.001440162056544&0.000798869250342&-0.00808541701429\\
12&&0.001464201727649&-0.0003416423396&0.00459909928294\\
13&&0.001744703197231&-0.000212938810296&0.006913583983173\\
14&&0.001100010416802&0.000223069933552&-0.00293630357091\\
15&&0.001061273315343&4.66867175896627E-06&-0.006226408282977\\
16&&0.001462284817869&-0.000101546964095&0.001484817717649\\
17&&0.001304160305372&3.77065191741915E-05&0.006024060541615\\
18&&0.001179839161665&3.78654896794631E-05&-0.00071186404194\\
19&&0.000350620499287&-2.25876364328412E-05&-0.006559629225319\\
20&&0.004041162071268&-8.36015569473183E-06&0.003989414923715
\end{tabular}
\caption{Coefficients of the data-driven entanglement witness used on Fig.~3 of the main text (tranverse-field Ising chain for $g=0.5$ and $T/J=0.28$).}
\label{tab_dd_Ising}
\end{table*}

\end{document}